\documentclass{elsart}
\usepackage{amssymb,amsfonts,amsmath}
\usepackage{epsfig}
\usepackage{newlfont}
\usepackage{graphicx}

\DeclareMathAlphabet{\mathitbf}{T1}{cmr}{bx}{it}

\newcommand{\s}{{\mathrm{s}}}

\journal{Computer Physics Communications}

\begin{document}
\begin{frontmatter}

\title{Simulating spin systems on IANUS, an FPGA-based computer}
\author[ferrara,infnfe]   {F.~Belletti},
\author[unizar,bifi]      {M.~Cotallo},
\author[unizar,bifi]      {A.~Cruz},
\author[ucm,bifi]         {L.~A.~Fern\'andez},
\author[unex,bifi]        {A.~Gordillo},
\author[ferrara,bifi]     {A.~Maiorano},
\author[ferrara,infnfe]   {F.~Mantovani\corauthref{cor1}},
\author[roma]             {E.~Marinari},
\author[ucm,bifi]         {V.~Mart\'\i{}n-Mayor},
\author[ucm,bifi]         {A.~Mu\~noz-Sudupe},
\author[ing,i3a]          {D.~Navarro},
\author[unizar,bifi]      {S.~P\'erez-Gaviro},
\author[unex,bifi]        {J.J.~Ruiz-Lorenzo},
\author[ferrara,infnfe]   {S.~F.~Schifano},
\author[unizar,bifi]      {D.~Sciretti},
\author[unizar,bifi]      {A.~Taranc\'on},
\author[ferrara,infnfe]   {R.~Tripiccione},
\author[unizar,bifi]      {J.~L.~Velasco}

\corauth[cor1]{Corresponding author. Filippo Mantovani (filimanto@fe.infn.it),
Dipartimento di Fisica, Universit\`a di Ferrara, via Saragat 1, I-44100 Ferrara (Italy), +39~0532~974610.}

\address[ferrara] {Dipartimento di Fisica, Universit\`a di Ferrara, I-44100 Ferrara (Italy)}
\address[infnfe]  {INFN, Sezione di Ferrara, I-44100 Ferrara (Italy)}
\address[unizar]  {Departamento de F\'{\i}sica Te\'orica, Facultad de Ciencias,\\ Universidad de Zaragoza, 50009 Zaragoza (Spain)}
\address[bifi]    {Instituto de Biocomputaci\'on y  F\'{\i}sica de Sistemas Complejos (BIFI), 50009 Zaragoza (Spain)}
\address[ucm]     {Departamento de F\'{\i}sica Te\'orica, Facultad de Ciencias F\'{\i}sicas,\\Universidad Complutense, 28040 Madrid (Spain)}
\address[unex]    {Departamento de F\'{\i}sica, Facultad de Ciencia, \\ Universidad de Extremadura, 06071, Badajoz (Spain)}
\address[roma]    {Dipartimento di Fisica, Universit\`a di Roma ``La Sapienza'', I-00100 Roma (Italy)}
\address[ing]     {Departamento de Ingenieria Electr\'onica y Comunicaciones,\\Universidad de Zaragoza, CPS, Maria de Luna 1, 50018 Zaragoza (Spain)}
\address[i3a]     {Instituto de Investigaci\'on en Ingenieria de Arag\'on ( I3A),\\Universidad de Zaragoza, Maria de Luna 3, 50018 Zaragoza (Spain)}


\begin{abstract}
We describe the hardwired implementation of algorithms for Monte Carlo
simulations of a large class of spin models. We have implemented
these algorithms as VHDL codes and we have mapped them onto a
dedicated processor based on a large FPGA device. The measured
performance on one such processor is comparable to $O(100)$
carefully programmed high-end PCs: it turns out to be even better for
some selected spin models. We describe here codes that we  are currently
executing on the IANUS massively parallel FPGA-based system.
\end{abstract}
\end{frontmatter}


\section{Introduction}

Numerical simulations with Monte Carlo (MC)
techniques of spin systems that show a complex behavior
(as, for example, because of the presence of frustrated quenched
disorder, the so called \textit{spin glasses})
require huge computational efforts: the non-trivial
structure of the energy-landscape, the long decorrelation time of the
dynamics, the need to analyze several different realizations of the
system all conspire to make the problem very challenging to clarify
numerically.  Reference \cite{sg} gives an introduction to numerical
spin glass systems, and discusses and elucidates a number of relevant
details.

One of the bottom lines is that traditional computers are not
optimized towards the computational tasks that are relevant in a
context of discrete variables: a large part of the needed CPU time is
spent essentially performing logical operations on individual bits or on
variables that can only appear in a few states,
at variance with arithmetics on
long data words (32 or 64 bits) which is the typical workload for
which computers are optimized today. This problem can be turned into
an opportunity by the proposal to develop a dedicated computer
optimized to handle the typical workload associated to these
applications. The use of Field Programmable Gate Arrays (FPGAs) adds
flexibility to a dedicated architecture: an FPGA based
system can be configured on-demand to perform with potentially very
high efficiency on a variety of different algorithms.

The FPGA approach for the simulation of spin systems has been proposed several
years ago \cite{sue}, and is now revisited in the IANUS
project, a massively parallel modular system based on a building block
of 16 high-performance FPGAs. The IANUS architectural concept has
been described in \cite{ianus}, while details of the hardware
prototype, currently undergoing final tests,  will be described
elsewhere \cite{hardPaper}. In this paper we focus on algorithm
mapping: we explore several avenues to map Monte Carlo algorithms
for spin systems on FPGAs, provide benchmark results for the
performance of several associated implementations, and present some
very preliminary results of large scale numerical simulations,
quantifying the potential performance of full-scale IANUS systems.

This paper is structured as follows: Section 2 describes the spin
models and the algorithms we have implemented as our first application
for IANUS. Section 3 gives details about the FPGA-based implementation
of those models and algorithms, covering various aspects of the VHDL
design. In section 4 we present some results and
performance figures for the test simulations of two different spin models.
Section 5 draws the conclusions of the work developed so far, and
outlines prospects for the near future.

\section{Monte Carlo simulations of Spin Glass systems}

\subsection{Models}\label{MODELSECT}

IANUS has been designed as a multipurpose reprogrammable computer; its
first application is the simulation of spin models. We are interested
in discrete models whose variables (the \textit{spins}) sit at the
vertexes of a $D-$dimensional lattice (the \textit{sites} of the
system). The spin variable associated to site $i$ ($s_i$) take
only a discrete and finite set of values (in some cases, just two
values).

We define an energy or cost function (the Hamiltonian $H$)
that drives the dynamics of the system. {\it Configurations}
of the system that appear in the course of the dynamics, once reached
an equilibrium state, are distributed according to the probability
function
\begin{equation}
\label{equ-p}
P \sim e^{-\beta H}\;,
\end{equation}
where $\beta$ is the \textit{inverse of the temperature} $T$ and
tunes the features of the type of configurations that appear at
equilibrium: when $\beta$ becomes large only configurations that
minimize $H$ are important (when $\beta\to\infty$ one looks for
optimal configurations, i.e. minima of $H$), while when $\beta\to 0$
the weight is not important and spin configurations become
equiprobable. Our local dynamics will allow us, in this way, to
determine important features of physical systems or for example, in
very strict analogy with it, of sets of equalities we want to
satisfy.

Each spin only interacts with its nearest neighbors, i.e. with spins
sitting at sites that are exactly one lattice spacing far away. The
strength of the interaction of spins $s_i$ and $s_j$ is
proportional to the value of a coupling $J_{ij}$, which in some models
(the classical \textit{Ising model}) is constant over all the
\textit{bonds} of the lattice (i.e. the connections among two first
neighboring sites), or can vary randomly from pair to pair (in this
case, for a given realization of the model, $J_{ij}$ depends on $i$ and
$j$: it is fixed when defining the realization of the model and does
not change during the dynamics). The model can be extended by adding
an external magnetic field $h_i$ at every site ($h_i$ can also be a
random variable), or also by considering the case of a diluted lattice
(only certain sites of the lattice are occupied by spins, while the
others are empty, depending on the value of the \textit{dilution}, 
$x_i=0,1$).

A generic Hamiltonian for two-state ($s_i=\pm1$) models has the form
\begin{equation}
\label{H_ising}
H - \sum_{<i,j>} J_{ij} x_i x_j s_i s_j
- \sum_i h_i x_i s_i\;,
\end{equation}
where $<i,j>$ means that the sum is taken on all pairs of
neighboring sites of the lattice.

Hamiltonians of the form (\ref{H_ising}) define several very
interesting models.  For instance, the \textit{Edward-Anderson} (EA)
spin glass \cite{ea} has $x_i=1$ and $h_i=0$ for all sites $i$, while
$J_{ij}$ takes random values ($\pm1$ in our work) with both positive
and negative support. The \textit{random field Ising model} (RFIM)
\cite{rfim,rfim_young} has $x_i=1$ and $J_{ij}=1$ everywhere, but the
field at each site takes random values $h_i=\pm\left| h \right|
$. Another interesting case is the \textit{diluted antiferromagnet in
a field} (DAFF) \cite{daff}, that has $J_{ij}=-1$ and $h_i=h$
everywhere while dilution $x_i$ takes randomly the value 0 or 1.

Models with two-state variables associated to the Hamiltonian
(\ref{H_ising}) are usually referred to as Ising-like and their
implementation on our FPGA-based computer are extensively discussed in
this paper.  Many other different spin models are very important: they
have for example higher space dimensionality or are defined on non
regular random graphs, longer range interactions or multivalued spin
variables (for example the Potts models).
In this note we also discuss the implementation of the dynamics of
a four-state \textit{glassy Potts model} \cite{potts},
defined by the Hamiltonian
\begin{equation}\label{H_potts}
H=-\sum_{<i,j>}\delta_{s_i \pi_{i,j}(\s_j)}\;,
\end{equation}
where the sum runs over first-neighbor sites, and the site variables
$s_i$ can take four values. $\pi_{i,j}$ are quenched random
permutations of $(0,1,2,3)$ (there are $4!$ of them): the pair of
first neighboring spins $(s_i,s_j)$ has non zero energy only if $s_i =\pi_{i,j}(s_j)$. This model
displays a number of features that are
typical of structural glasses, and could hopefully help
describe the glassy state, that stays difficult to understand.

\subsection{Algorithms}

Our goal is to analyze, by numerical Monte Carlo simulations,
the properties of the models described above.
We have implemented for the IANUS processor two well-known
algorithms, namely \textit{Metropolis} and \textit{Heat Bath}.

Both algorithms update a single spin at a time: they sweep the
entire lattice and then start again. After a (long enough) number of
steps one reaches, as discussed before, an equilibrium state,
and the spin configurations that appear during the dynamics are
typical of the probability distribution (\ref{equ-p}).

In
the case of the Metropolis algorithm
we propose to update a spin $s_i$, and
we calculate the corresponding energy change $\Delta E$.
If $\Delta E < 0$, the update makes the energy function lower, and
change is accepted. Otherwise
we do not necessarily refuse the update
(this would be a $\beta=\infty$ dynamics,
where we move to the closest local minimum of $H$)
but we accept it with a probability
$p_M=\exp(-\beta \Delta E)$.

In the case of the Heat Bath algorithm
we directly select the new value of the spin
with a probability
proportional to the Boltzmann factor
\begin{equation}
  P_{HB}(s_i = +1)  = \frac{e^{-\beta E_+}}{e^{-\beta E_+}+e^{-\beta E_-}}\;,
\end{equation}
where $E_+$ and $E_-$ are the local energies of the two spin
configurations
for spin $s_i$ pointing up
($s_i=+1$) or down
($s_i=-1$), respectively. Since when we change $s_i$
only a few terms of the energy function change (the ones containing
spin $s_i$ and its first neighbors), this is a fast and easy computation.

We define one full MC sweep to be the iteration of these simple
steps for all sites of the lattice.  The spin configurations
that appear during the dynamical process we are simulating are
\textit{correlated}: a spin configuration depends on the ones that
appeared at former times, and only when we consider large time
separation among two such configuration we can consider them as
independent. In this way we can define a \textit{correlation time}
(that depends on $\beta$ and characterizes the dynamics), that we can
roughly define as the number of Monte Carlo sweeps it takes to make
two spin configurations uncorrelated (see refs. \cite{sokal} and
\cite{amit_mayor}).  An estimate of this \textit{correlation time} is
usually calculated during the simulation, taking configurations at
various times and measuring their correlation.

Other algorithms are used in some simulations, as they offer higher
efficiency in decorrelating the spin configurations (see
\cite{algorithms} for a review). On one side no very effective
specialized algorithm exist, for example, for the very interesting
case of spin glasses (we have in mind here mainly cluster
algorithms), and their implementation on IANUS would probably  not be
very effective: so we do not use this kind of algorithms, and stay
with simple, local dynamics. On the other side, algorithms like
Parallel Tempering \cite{PT} are crucial for simulating complex
systems like spin glasses, but their implementation on our FPGA
based devices is a trivial add-on so we do not discuss
them here.

\section{Hardware implementation}\label{VHDLSECT}

\subsection{Parallelism}\label{PARALLSECT}

The guiding line of our implementation strategy is to try to express all
the parallelization opportunities allowed by the FPGA architecture,
matching as much as possible the potential for parallelism offered by
spin systems.  Let us start by noticing that, because of the locality
of the spatial interaction \cite{algorithms}, the lattice can be split
in two halves in a checkerboard scheme (we are dealing with a so-called
\textit{bipartite lattice}),  allowing in principle the parallel
update of all white (or black) sites at once. Additionally, one can
further boost performance by updating in parallel more copies of the
system. We do so by updating at the same time two spin lattices (see
later for further comments on this point).  Standard PCs cannot
efficiently exploit all available parallelism for several reasons, the
most fundamental one being memory architecture, that prevents the
processor from gathering fast enough all variables associated to the
computation. Sharing the simulation between several computers is an
interesting parallel solution, but optimization has a bottleneck in
the limited bandwidth and large latency associated to communication
patterns (see \cite{ianus}).

The hardware structure of FPGAs allows exploitation of the full
parallelism available in the algorithm, with the only limit of logic
resources. As we explain below,
the FPGAs that we use (Virtex4/LX160 and Virtex4/LX200,
manufactured by Xilinx) have enough resources for the simultaneous
update of half the sites for lattices of up to $8^3$ sites. For larger
systems there are not enough logic resources to generate all the
random numbers needed by the algorithm (one number per update, see
below for details), so we need more than one clock cycle to update the
whole lattice. In other words, we are in the very rewarding situation
in which: i) the algorithm offers a large degree of allowed
parallelism, ii) the processor architecture does not introduce any
bottleneck to the actual exploitation of the available parallelism,
iii) performance of the actual implementation is only limited by the
hardware resources contained in the FPGAs.

We have developed a parallel update scheme, supporting 3-D lattices
with $L \geq 16$, associated to the Hamiltonian of (\ref{H_ising}).
One only has to tune a few parameters to adjust the lattice size and
the physical parameters defined in $H$. We regard this as an important
first step in the direction of creating flexible enough libraries of
application codes for an FPGA-based computers.

The number of allowed parallel updates depends on the number of logic
cells availables in the FPGAs. For the Ising-like codes
developed so far, we update up to $1024$ sites per clock cycle on a Xilinx
Virtex4-LX200, and up to $512$ sites/cycle for the Xilinx
Virtex4-LX160.  The algorithm for the Potts model requires more logic
resources and larger memories, so performances lowers to $256$ updates/cycle on
both the LX200 and LX160 FPGAs.

\subsection{Algorithm Implementation}\label{MEMSECT}
\begin{figure}[h]
\begin{center}
\includegraphics[width=10cm]{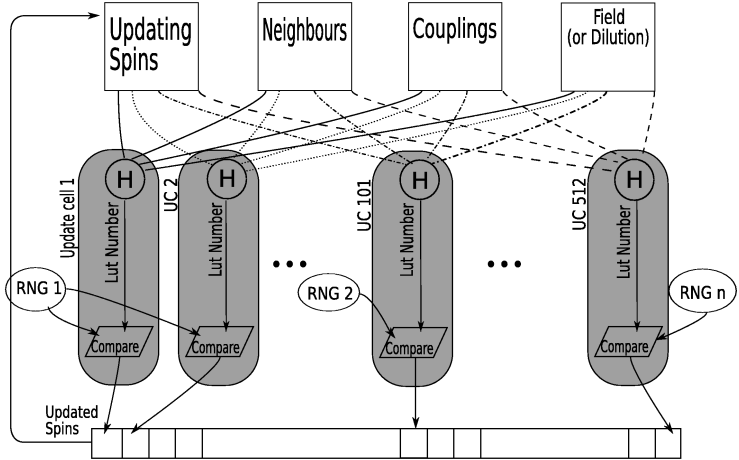}
\caption{Parallel update scheme. The spins that must be updated, their neighbors,
the couplings and all other relevant values are passed to the update
cells where the energy is computed. The result is used as a pointer to
a Look-up Table (LUT).  The associated value is compared with a random
number (RNG), and following the comparison, the updated spin value is
computed.}
\label{fig:update}
\end{center}
\end{figure}

We now come to the description of the actual algorithmic architecture,
shown in fig. \ref{fig:update}.

In short, we have a set of update cells ($512$ in the picture): they
receive in input all the variables and the parameters needed 
to perform all required arithmetic and logic
operations, and compute the updated value of the spin variable. Data
(variables and parameters) are kept in memories and are fed to the
appropriate update cell.  Updated values are written back to memory,
to be used for subsequent updates.

The choice of an appropriate storage structure for data and the
provision of enough data channels to feed all update cells with the
data they need is a complex challenge; designing the update cells is a
comparatively minor task. Hence we describe first the memory
structures of our codes, followed by some details on
the architecture of the update cells.

Virtex-4 FPGAs have several small RAM-blocks that can be grouped
together to form bigger memories. We use these blocks to store all
data items: spins, couplings, dilutions and external fields. The
configurable logic blocks are used for random number generators and
update cells.

To update one spin of a three dimensional model we need to read its
six nearest neighbors, six couplings, the old spin value (for the
Metropolis algorithm) and some model-dependent information such as the
magnetic field for RFIM and the dilution for DAFF. All these
data items must be moved to the appropriate update cells, in spite of
the hardware bottleneck that only two memory locations in each block
can be read/written at each clock cycle.

\begin{figure}[h]
\begin{center}
\includegraphics[width=10cm]{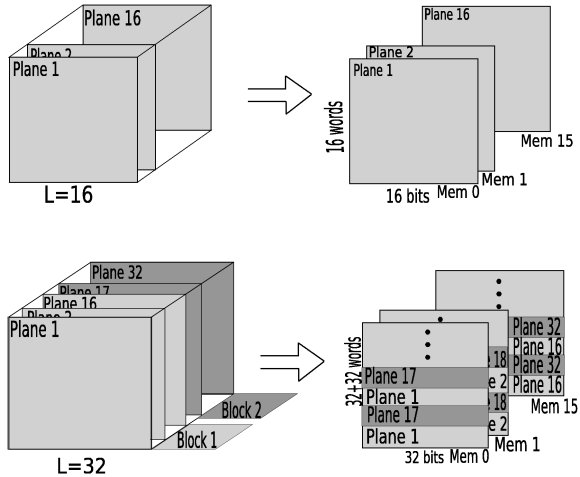}
\caption{Examples of the spin memory structure: L=16 and L=32.}\label{mem}
\end{center}
\end{figure}

Let us analyze first the Ising models, considering for definiteness
$L=16$.  We choose to use an independent memory of size $L^3$ for each
variable.  This is actually divided into smaller memories, arranged so
that reading one word from each gives us all the data needed for a
single update cycle.  We need $16^3=4096$ bits to store all the spins
of one replica. We have $16$ vertical planes, and save each plane in a
different memory of width $16$ bits and height $16$ (see
Fig.\ref{mem}).  In this simple case the logic resources within the FPGA allow to 
update one whole horizontal plane in one clock cycle (because we mix
the two bipartite sublattices of two different copies of the system,
see the following discussion), and the reading rate matches
requirements, as we need to read only one word from each of the
sixteen memories.

The configuration is slightly more complex when the size of the
lattice grows and the update of a full plane in just one clock cycle
is no longer possible. In this case we split each plane in a variable
number of blocks $N_B$, adjusted so that all the spins of each block
can be updated in one clock cycle. The number of independent memories is
$L / N_B$, as only these need to be read at the same time. The
data word still have width $L$, while the height is $L \times N_B$ to
compensate for the reduced number of memories. Considering $L=32$, for
example, we have a plane made of $32^2=1024$ spins, too large to be
updated in one cycle (in the Xilinx Virtex4-LX160). We split it in two
blocks of $32 \times 16=512$ spins each. To read $16$ lines every
clock cycle we store the spins into $16$ memories, each of width $32$
bits and height $32 \times 2$: the total size of the memory is still
$32^3$ bits.

\begin{figure}[h]
\begin{center}
\includegraphics[width=10cm]{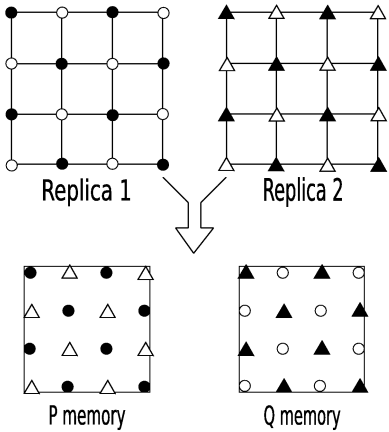}
\caption{Structure of spin configuration memories: meshing of replicas.}\label{fig:12_PQ}
\end{center}
\end{figure}

As already remarked, we simulate two different replicas in the same
FPGA. This trick bypasses the parallelism limit of our MC algorithms
(nearest neighbors cannot be updated at the same time, see
\cite{ianus} ). We mesh the spins of the two replicas in a way that
puts all the whites of one replica and the blacks of the other in
distinct memories that we call respectively $P$ and $Q$ (see
Fig.\ref{fig:12_PQ}). Every time we update one slice of $P$ we handle
one slice of whites for replica $1$ and one slice of blacks for
replica $2$. Obviously the corresponding slice of memory $Q$ contains
all the black neighbors of replica $1$ and all the white neighbors
of replica $2$.

The amount of memory available in the FPGA limits the lattice
size we can simulate and the models we can implement. In both the
Virtex4-LX160 and LX200 it is possible to simulate EA, RFIM and DAFF
models in $3D$ with size up to $L=88$ (not all smaller sizes are
allowed). Because of the dramatic
\textit{critical slowing down} of the dynamics of interesting complex
spin models these size limits are confortably larger of what we can
expect to be able study (even with the tremendous power made
available by IANUS) in a reasonable amount of (wall-clock) time:
memory size is presently not a bottle-neck.

Things are even more complicated when one considers multi-state
variables, as more bits are required to store the state of the system
and all associated parameters.  In the four state Potts model (see
sec.  \ref{MODELSECT}) the site variables need two bits and the
couplings eight bits. In order to keep a memory structure similar to
that outlined before we store each bit in a different memory. For
example a lattice with $L=16$ requires $16 \times 2$ memories for the
site variables (they were sixteen in the Ising case), and $16 \times
8$ memories for the couplings.

The lattice meshing scheme is maintained. With our reference FPGAs we
can simulate three dimensional Potts model with at most $L=40$ and
four dimensional Potts model with $L=16$.

We now come to the description of the update cells. The Hamiltonian we
have written is homogeneous: the interaction has the same form for
every site of the lattice, and it only depends on the values of the
couplings, the fields and the dilutions. This means that we can write
a standard update cell and use it as a black box to update all sites:
it will give us the updated value of the spin (provided that we feed
the correct inputs). This choice makes it easy to write a parametric
program, where we instantiate the same update cell as many times as needed.

We have implemented two algorithms: Metropolis and Heat Bath. The
update cell receives in input couplings, nearest neighbors spins,
field and dilution and, if appropriate, the old spin value (for the
Metropolis dynamics). The cell uses these values as specified by 
(\ref{H_ising}) and computes a numerical value between $0$ and $15$
(the range varies depending on the model) used as an input to a
LUT. The value read from the LUT is compared with a random number and
the new spin state is chosen depending on the result of the
comparison. Once again, things are slightly different for the Potts
model due to the multi-state variables and couplings.

Our goal is to update in parallel as many variables as possible,
which means that we want to maximize the number of cells that will
be accessing the LUT at the same time. In order to avoid routing
congestion at the hardware layer we replicate the LUTs: each
instance is read only by two update cells. The waste in logic
resources -- the same information is replicated many times within
the processor -- is compensated by the higher allowed clock
frequency.

\subsection{Random numbers}

Monte Carlo methods depend strongly on the random numbers used to
drive the updates: this determines the imperative need to implement a
very reliable pseudo-random number generator (RNG), that produces a
sequence of numbers under the selected distribution, with no known or
evident pathologies.

We use the Parisi-Rapuano shift register method
\cite{rapuano} defined by the rules:
\begin{eqnarray}
\label{eq:rng_rules}
I(k) & = & I(k-24)+I(k-55) \\
R(k) & = & I(k)\otimes I(k-61)\;, \nonumber
\end{eqnarray}
where $I(k-24)$, $I(k-55)$ and $I(k-61)$ are elements (32-bit wide)
of a so called
\textit{wheel} that we initialize with externally generated random values.
$I(k)$ is the new element of the updated wheel,
and $R(k)$ is the generated pseudo-random value.

\begin{figure}[h]
\begin{center}
\includegraphics[width=10cm]{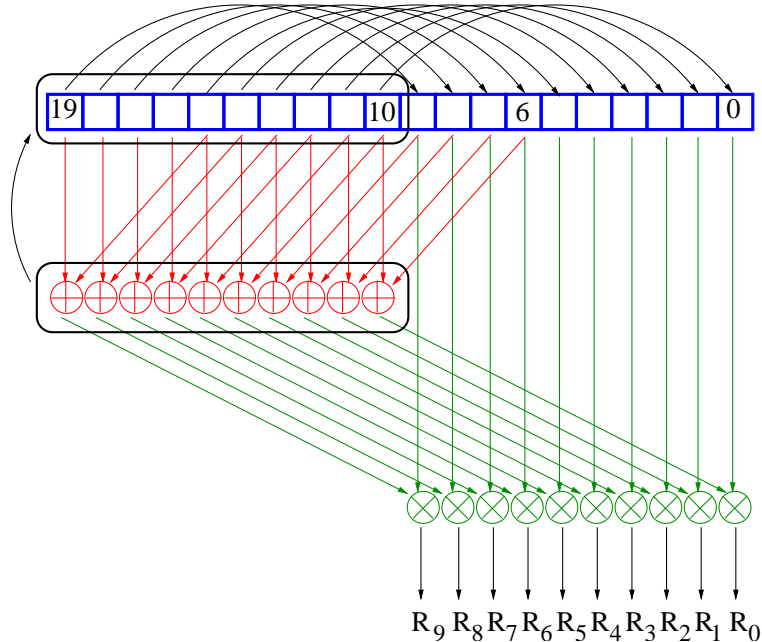}
\caption{Hardware implementation of the Parisi-Rapuano RNG. For
graphical reasons the example refers to a wheel of only 20 numbers and
following the reduced equations $I(k)=I(k-10)+I(k-14)$ and
$R(k)=I(k)\otimes I(k-20)$. The combinatorial logic complexity grows
when producing more numbers.}\label{fig:multi_rng}
\end{center}
\end{figure}

A straightforward implementation of this algorithm produces one random
number at each step, for each wheel that we maintain. A wheel uses
many hardware resources (in our case we use the three pointer
values $24$, $55$ and $61$ so we need to store $62$ numbers), and the
random number generator is a system bottleneck, since the number of
updates per clock cycle is limited by how many random values we are
able to produce.  A large performance improvement comes from the
implementation of the wheel through logic (as opposed to
memory) blocks, as the former can be written in cascade-structured
combinatorial logic that may be tuned to produce several numbers per
clock cycle.  We can exploit this feature and use a limited number of
wheels to produce more numbers, thus increasing the number of updates
per clock.  Remember that to produce one random number we must save
the result of the sum of two values and then perform the XOR with a
third value.  The wheel is then shifted and the computed sum fills the
empty position. All this is done with combinatorial logic, so one can
produce various pseudo-random numbers simply replicating these
operations and, of course, increasing logic complexity.  A schematic
representation of a simplified case is given in
fig. \ref{fig:multi_rng}.

The logic complexity of the implementation depends  on the
parameters of (\ref{eq:rng_rules}) and on the quantity of random
numbers we need. We use one wheel to generate up to $96$ numbers per
clock (so more wheels are active at the same time to compute all
needed random values).

To keep the wheel safely below its period limit we choose to reload the
wheel every now and then (for example every $10^7$ MC sweeps).

With respect to the choice of $32$-bit random numbers, we have
verified that this word size is sufficient for the models we want to
simulate (our tests show that $24$-bit would be enough). Other models
may require \textit{better} random numbers. We do not address this
issue here. We just note that generating random numbers of larger size
(e.g., $40$ or even $64$-bit) would be straightforward, at the price,
of course, of a larger resource usage.

All in all, our carefully handcrafted VHDL codes use a very large fraction of
the available FPGA resources, as measured by the number of used logic blocks 
and RAM-blocks. 
The following table shows figures
for the Ising-like and Potts-model codes. Mapping has been
performed with the ISE toolkit made available by Xilinx. The Ising-like code is
limited by logical resources, while the Potts model, with its larger storage
requirements, is limited by available memory space.

\begin{table}[ht]
\begin{center}
\begin{tabular}{|c|c|c|c|c|}
\hline
Model              & Resource    & Number used&  \% (LX160) & \% (LX200)\\
\hline
\hline
Ising-like         &  Log. blocks& $157,649$   & $(117 \%)$& $88 \%$ \\
(1024 updates)     &  RAM-blocks & $160$       &  $56 \%$  & $47 \%$ \\
\hline
Ising-like         &  Log. blocks& $83,651$    & $62 \%$   & $46 \%$ \\
(512 updates)      &  RAM-blocks & $80$        & $28 \%$   & $23 \%$ \\
\hline
Potts $q=4$        &  Log. blocks& $117,586$   & $86 \%$   & $66 \%$ \\
(256 updates)      &  Ram-blocks & $224$       & $77 \%$   & $67 \%$ \\
\hline
\end{tabular}

\caption{Use of FPGA resources, as absolute values and as fraction of available
blocks on our FPGAs, for the Ising-like and Potts codes.
In both cases, the 3-D lattice has a linear size $L = 32$. The Ising-like code
is limited by available logic resources , while the Potts code is
memory-limited.}
\label{tab:hardware}
\end{center}
\end{table}

\section{Benchmark tests}

\subsection{Edward-Anderson spin glass model}

We have simulated an $L=32$ $3D$ system at $\beta(=1/T)=0.878$. The
number of MC sweeps sums up to $8\times10^9$.
See reference \cite{sue_sg} for
previous
simulations done with the special purpose machine SUE
on a  lattice of size $L=20$

\begin{figure}[h]
\begin{center}
\includegraphics[angle=270,width=10cm]{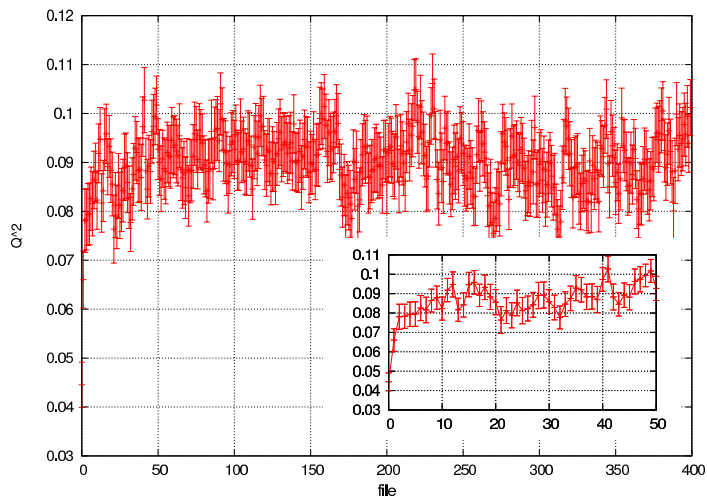}
\caption{Evolution of $q^2$: the {\it x}-axis scale is $10^6$ MC sweeps
per file}\label{fig:Q2hist}
\end{center}
\end{figure}

Checking that thermalization has been reached is a common and
non-trivial problem in spin glass simulations. Here we provide only
a short review of our analysis:
full details will be published elsewhere. In our
early tests, configurations were copied to the host computer every
$10^6$ MC sweeps, because, when performing these tests, we had a
very slow communication channel to the host
\footnote{The situation has now improved dramatically.
The I/O interface to the host computer is
discussed in details in \cite{hardPaper}}.
This value is too high to see clearly
the evolution towards equilibrium along the
first sweeps. Fig. \ref{fig:Q2hist} shows the MC history of a
physically meaningful quantity, the squared overlap $q^2$; a zoom of
the leftmost part of the plot (inset graphic) shows the drift from
the initial value ($0.045$) to a value probably very close to
the equilibrium value in less then $50 \times 10^6$ sweeps.

\begin{figure}[h]
\begin{center}
\includegraphics[angle=270,width=10cm]{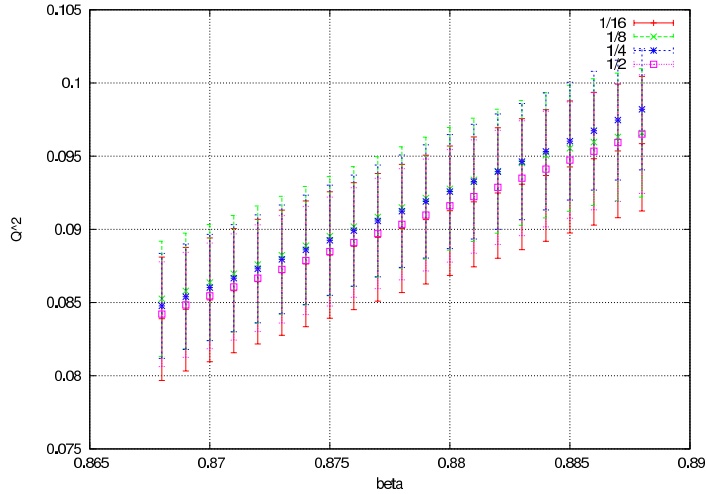}
\caption{Thermalization test of the squared overlap $q^2$ as a
function of $\beta=1/T$.}\label{fig:Q2therm}
\end{center}
\end{figure}

We have analyzed the  thermalization rate
also with the standard $log_2$L data
binning: we divide the data points into four groups of variable size
(namely the last half of the measures, then the previous quarter,
the previous eighth and the sixteenth before this) and then average
over all samples in each group. From the smaller $1/16$th to the
bigger $1/2$ group the averaged value is expected to shift toward
its equilibrium value.
Fig. \ref{fig:Q2therm} shows the behavior of the squared overlap $q^2$.
The time dependence we observe on the latest data is very small, it
does not expose any systematic drift and is
surely far smaller than the statistical error.

\begin{figure}[h]
\begin{center}
\includegraphics[angle=270,width=10cm]{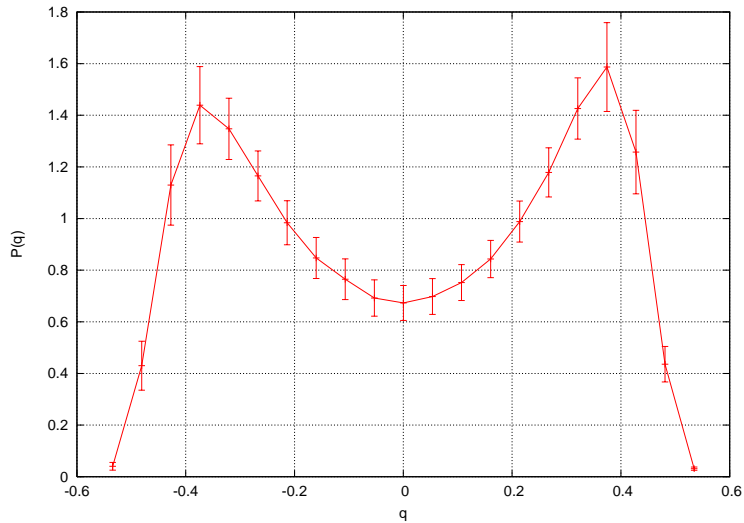}
\caption{Distribution of the overlap $q$, showing a reasonably
symmetric behavior, within error bars.}
\label{fig:Pq}
\end{center}
\end{figure}

A clean visual representation of the system thermalization is also given via
the average overlap probability distribution $P(q)$, which should be
symmetric at equilibrium (with no external field),
as shown in fig. \ref{fig:Pq}: this is obviously only a necessary
condition for thermalization, but it surely is a good sign.

\subsubsection{Performance}

The algorithms described in the previous section are mapped on the
selected FPGAs with a system clock of 62.5 MHz. At each clock cycle,
$512$ ($1024$) spins are updated on an LX160 (LX200), corresponding to
an average update time of $32$ ps/spin ($16$ ps/spin).

It is interesting to compare these figures with those appropriate for
a PC.  Understanding what exactly has to be compared is not completely
trivial.  The fastest PC code for spin model simulations available to
us is the multi-spin coding, which updates in parallel a large number
of (up to 128) samples of the system at the same time, using only one
random number generator, which is shared across all samples: this
scheme is useful to obtain a large number of configuration data,
appropriate for statistical analysis. We call this an asynchronous
multi-spin coding (AMSC) as inside each sample there is no
parallelism.

As we have stated before, the biggest problem with the models we want
to study is the decorrelation time, and the large number of Monte
Carlo sweeps that it may take to bring a configuration to
equilibrium. The AMSC procedure has a serious problem here since each
sample evolves for the same number of sweeps as if it were the only
one being simulated. In other words, efficient codes on a PC achieve
high overall performance by simulating for relatively \textit{short}
MC time a \textit{large} number of independent samples.  A code that
updates in parallel more spins belonging to the \textit{same} system
would be more useful to attain equilibrium, when working on large
systems. The resulting algorithm, synchronous MSC (SMSC), takes less
time to simulate one sample, but the global performance is lowered
because of more complex operations involved and the need to use an
independent random number for each spin. The SMSC PC-code available to us
updates up to 128 spins of a single sample. Synchronous codes are not
commonly used in PC based numerical simulations because of their
globally poor performances.

Generally speaking we think that comparison with a SMSC code is
appropriate for a single FPGA system, while comparison with an AMSC
code is more relevant when considering a massively parallel IANUS
system (we plan to build a system with 256 FPGA-based nodes). Here
we simply present our preliminary comparison data for both cases in
table \ref{tab:ea_comp}.

\begin{table}[ht]
\begin{center}
\begin{tabular}{|c|c|c|c|c|}
\hline
& LX160 & LX200 & PC (SMSC) & PC (AMSC)\\
\hline
\hline
Update Rate
& 32 ps/spin & 16 ps/spin& 3000 ps/spin& 700 ps/spin\\
\hline
\end{tabular}

\caption{Comparing the performances of two Xilinx Virtex4 FPGAs and two
         different codes running on a high-end PC.}
\label{tab:ea_comp}
\end{center}
\end{table}

The MSC values are referred to an Intel Core2Duo (64 bit) 1.6 GHz
processor. Inspection of table \ref{tab:ea_comp} tells us that one
LX160 runs as fast as 90 PCs, while the LX200 performance is
comparable to that of 180 PCs. In other words, the $8\times 10^9$
MC iterations required to thermalize a lattice of size $L=32$ took
approximately $6$ hours to be completed on just one Virtex4-LX200: they
would take $18$ days on a PC running the SMSC algorithm.

Performance comparison with published work is difficult. As far as
we know the SMSC is not used for massive simulations, so data on the
performances of this algorithm is not widespread. The AMSC is
commonly used. Even if it is considered almost a standard in spin
glass simulations, we have not been able to find recent speed
analysis. The seminal works on this algorithm \cite{MSC_heuer} and
\cite{MSC} are way too old in technology terms to allow a fair
comparison. All in all, we believe that the codes that we have used for
performance comparison are
state-of-the-art PC implementations, and further optimizations
could at most improve the performances by some $10 - 20 \%$; we
conclude that the performance of one FPGA-processor is roughly
$\simeq 100$ times better than possible on a PC.

\subsection{Potts model}

Simulations for the three and four dimensional $L=16$ four-state Potts
model have been run at various temperatures. Previous works on the
$4D$ model \cite{potts} could only thermalize a lattice of size $L=5$,
and study $L=8$ in the warm phase. To obtain good results with out
tests we had to run long simulations, which took up to $18\times
10^{11}$ MC iterations for the $3D$ case ($\beta=2.4$) and $6\times
10^{10}$ in $4D$ ($\beta=1.405$).

\begin{figure}[h]
\begin{center}
\includegraphics[angle=270,width=10cm]{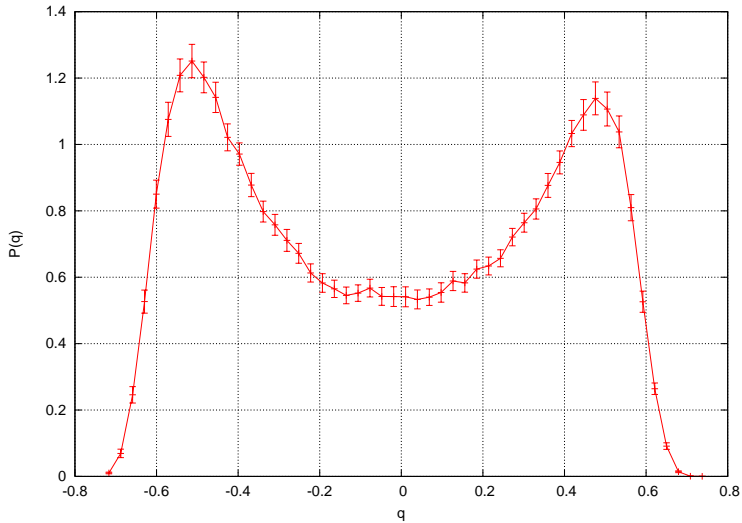}
\caption{Distribution of the overlap $q$ for the four-states three
dimensional Potts model with $L$ = 16 at $\beta = 2.4$.
}\label{fig:Pq_potts}
\end{center}
\end{figure}

Also in this case the thermalization has been analyzed using
$log_2$ data binning.
Fig. \ref{fig:Pq_potts} shows the average $P(q)$ for the Potts
disordered model. The symmetry is not as good as for the Ising spin
glass.

\subsubsection{Performance}

The update algorithm of the Potts model requires more complicated operations.
We do not have an MSC version for this case but we do not expect great
improvements compared with the non-MSC code we have used: the size of the
spin variables (four bits) and the structure of the update algorithm do not
leave as much space for parallelization and high performances on a PC
as it was the case for the EA model.
We expect that an optimized code would be at most twice faster (halving the
times shown here).

Our FPGA-based implementation does not suffer the increased complexity
as much as the PC generic architecture: apart from the smaller sizes permitted,
the number of sites that can be updated at the same time reduces
only by a factor four with respect to EA.

\begin{table}[ht]
\begin{center}
\begin{tabular}{|c|c|c|c|}
\hline
& LX160 & LX200 & PC\\
\hline
\hline
Potts 3D
& 125 ps/spin & 64 ps/spin& 117 ns/spin\\
\hline
Potts 4D
& 125 ps/spin & 64 ps/spin& 150 ns/spin\\
\hline
\end{tabular}
\caption{Potts model performances in two Xilinx Virtex4 FPGAs and in a PC.}
\label{tab:potts_comp}
\end{center}
\end{table}
Table \ref{tab:potts_comp} shows the comparison with an Intel Pentium IV 3.2 GHz processor.
Simulating a three dimensional lattice, the smaller LX160 performs as $900$ PCs
approximately, while the LX200 works as fast as $1800$ PCs. In $4D$-models
these numbers change respectively to 1200 for LX160 and 2300 for
the LX200.

Once again we point out that these results refer to just one FPGA.
A IANUS computer has $16$ FPGA devices per board, improving, with a (needed)
embarrassing parallelism performance increased by a factor $16$: a
complete IANUS computer will probably have $16$ boards, bringing this
factor to $256$ and the performance global ratio of a LX200 based
machine to a PC for a $4D$ disordered Potts model
to a number of the order of half a million.

\section{Conclusions}
This paper has described the implementation on a FPGA based
engine performing Monte Carlo simulation of some classes of spin models.
The main results of our work can be summarized as follows:

\begin{itemize}

\item The simulation engine exploits all the parallelization space in
principle available in the algorithm and its performance is limited
only by available hardware resources.

\item Measured performance are outstanding if compared with figures
available for traditional PCs: one FPGA has the same performance as $
\simeq 100$ PCs for the Edwards-Anderson spin glass model. Comparison
data is even more impressive for the Potts model, whether
the speed-up factor is close to $\simeq 1000$.

\item The FPGA-based processor has shown stable operation for extended
time periods: thanks to this reliable behavior we have been able to
collect a large number of spin configurations.

\end{itemize}

The work described here will continue with actual physics runs in
which large statistics will be collected for larger systems, as soon
as large IANUS systems are available. We also continue to work on
the development of simulation codes for other algorithms of
scientific interest. Work is in progress in such areas as random
graphs, surface growing and protein docking.

\section*{Acknowledgments}
The help of G. Poli in the development of the IANUS Ethernet interface
is warmly acknowledged.

\end{document}